\documentclass{webofc}
\usepackage[varg]{txfonts}   
\usepackage{xcolor}
\usepackage{todonotes}
\usepackage{placeins}
\usepackage{orcidlink}

\newcommand{\Sherpa}{S\protect\scalebox{0.8}{HERPA}\xspace}

\newcommand{\Amegic}{A\protect\scalebox{0.8}{MEGIC}\xspace}

\newcommand{\Comix}{C\protect\scalebox{0.8}{OMIX}\xspace}

\newcommand{\HepMC}{H\protect\scalebox{0.8}{EP}MC\xspace}

\makeatletter
\renewcommand{\@oddfoot}{\hfil\thepage\hfil}  
\renewcommand{\@evenfoot}{\hfil\thepage\hfil} 
\makeatother
\begin{document}
\title{Phase space sampling with Markov Chain Monte Carlo methods}
%
%

\author{
\firstname{Salvatore} \lastname{La Cagnina}\inst{1}\fnsep\orcidlink{0000-0002-7786-9199}\thanks{\email{salvatore.lacagnina@tu-dortmund.de}}
        \and
\firstname{Cornelius} 
\lastname{Grunwald}\inst{1}\fnsep\orcidlink{0000-0002-0588-8788}
\and
        \firstname{Timo} \lastname{Janßen}\inst{2,3}\fnsep\orcidlink{0000-0001-9466-477X} \and
        \firstname{Kevin} \lastname{Kröninger}\inst{1,4}\fnsep\orcidlink{0000-0001-9873-0228}
        \and
        \firstname{Steffen} \lastname{Schumann}\inst{2}\fnsep\orcidlink{0000-0003-0330-3990}
}

\institute{TU Dortmund University, Department of Physics, Dortmund, Germany
\and
           Institute for Theoretical Physics, University of G\"ottingen,
Germany
\and
Campus Institute Data Science, University of G\"ottingen, Germany
\and
TU Dortmund - Center for Data Science and Simulation, Dortmund, Germany
          }

\abstract{%
  We present a study on using Markov Chain Monte Carlo (MCMC) techniques to explore the high-dimensional and multi-modal phase space of scattering events at high-energy particle colliders. To this end, we combine the BAT.jl package that provides implementations of a variety of MCMC algorithms with the \Sherpa event generator framework. We discuss technical aspects of the implementation and the resulting algorithm and present first results for the process of $Z+3$ jets production at the LHC. 
}

\maketitle
\noindent \begin{center}
  Presented at: \\
  Conference on Computing in High Energy and Nuclear Physics (CHEP 2024) \\October 19--25, 2024, Krakow, Poland\\
  Submitted to EPJ Web of Conferences (EPJ WoC)
\end{center} 

\section{Introduction}

The precise simulation of scattering processes, such as those induced in collider experiments, is a central objective in particle physics. One successful example is the description of inelastic proton--proton ($pp$) scattering at the Large Hadron Collider (LHC) and its experiments. Processes with probabilities of occurrence differing by orders of magnitude are described with accuracies in the percentage range by corresponding simulation tools. While the LHC experiments already provide an unprecedented amount of data, the High Luminosity LHC (HL-LHC) -- starting operation in 2030 -- will exceed this by at least an order of magnitude during its entire lifetime. This poses a prominent challenge for the description of the data: firstly, the sheer amount of simulated data must increase by a similar factor as the experimental data, and secondly, collision events with even more complex final states than before will be observed and must also be addressed by the simulation. The last point is of particular importance when it comes to the search for new physical effects and the improved understanding of the Standard Model of particle physics. This results in the need for huge computing resources as documented, e.g., in the Snowmass 2021 process~\cite{Elvira:2022wyn,Campbell:2022qmc} and by the HEP Software Foundation (HSF) Physics Event Generator Working Group~\cite{HSFPhysicsEventGeneratorWG:2020gxw}.

Inelastic $pp$ scattering is usually described in a modular fashion starting with the hard scattering process, followed by parton-showering, hadronization and particle decays~\cite{Buckley:2011ms}. The result is a simulated set of stable particles as they would appear shortly after the scattering process. The comparison with experimental data requires a second simulation step: the stable particles are tracked through a (comparatively large) detector volume and the reaction of the detector elements is simulated. In all further processing steps, experimental and simulated data are treated equally. There are two critical aspects in the full simulation chain. One is the efficient generation of parton-level events with complex final states, e.g. $Z$-boson production with several additional jets. The other is the simulation of the full detector response. 
Our study focuses on the first aspect. The challenge in the generation of events is rooted in drawing multi-dimensional sets of random numbers from the fully differential cross section that describes the process under study, a procedure referred to as phase space sampling. Following Fermi's Golden rule, the shape of the differential cross section is dominated by the square of the matrix element, $\left| \mathcal{M} \right|^{2}$, that describes the transition from an initial to a final state. The expression $\left| \mathcal{M} \right|^{2}$ can have several narrow local modes and can vary in size over several orders of magnitude. Depending on the complexity of the expressions contributing, the evaluation of the matrix element can be computationally expensive compared to the other contributions to the differential cross section.
While most Monte Carlo (MC) generators use rather traditional sampling methods, e.g. importance sampling, several alternatives and variants are being developed. The approach presented in this study is to use Markov Chain Monte Carlo (MCMC) for phase space sampling. MCMC algorithms have been developed in the 1940ies to solve numerical problems in nuclear research. The most prominent example is the classical Metropolis algorithm~\cite{Metropolis:1953am}. The aim of our study is twofold: firstly, we demonstrate that MCMC algorithms can be used to sample the fully differential cross section of a process and produce the corresponding distributions, and  secondly, we study the potential for using MCMC algorithms for event generation. The study will be based on two concrete software tools, namely the \Sherpa event generator~\cite{Gleisberg:2008ta,Sherpa:2019gpd,Sherpa:2024mfk} and BAT.jl for its collection of MCMC algorithms~\cite{Schulz:2020ebm}.
In Sec.~\ref{sec:methods} we briefly review the basics of phase space sampling and MCMC methods, thereby introducing the methods used in \Sherpa and BAT.jl. In Sec.~\ref{sec:batsh}
we describe the novel interface between the two codes and present first results for the case of $Z+$jets production at the LHC. A summary and an outlook are given in Sec.~\ref{sec:conclusions}.
 
\section{Methods}
\label{sec:methods}
\subsection{Phase space sampling basics}

The evaluation of a hadronic cross section amounts to the integration of the contributing partonic matrix elements $\left| \mathcal{M}_{ab\to X_n} \right|^{2}$, convoluted with parton-density functions, over the considered $n$-particle final state phase space volume $\Omega$, i.e.,
\begin{align}
\sigma_{pp\to X_n}=\sum_{a,b}\int\limits_{\Omega} f_{a/p}(x_a,\mu_F)f_{b/p}(x_b,\mu_F)\left| \mathcal{M}_{ab\to X_n} \right|^{2}(\Phi_n,\mu_R,\mu_F)\, dx_a dx_b d\Phi_n\,.    
\end{align}
Considering massless final state particles, the Lorentz-invariant phase space element reads
\begin{align}
    d\Phi_n = \prod\limits_{i=1}^n\frac{d^3p_i}{(2\pi)^32E_i}\delta^{(4)}\left(p_a+p_b-\sum\limits_{k=1}^np_k\right)\,,
\end{align}
where the $\delta$-function ensures four-momentum conservation. Instead of sampling directly the 
particle momenta $\{p_i\}$, typically samples $\theta$ are drawn from a unit hypercube of dimension $D=3n-4$, employing some mapping $\Pi:[0,1]^D\rightarrow \Omega, \,\Pi(\theta)\mapsto \{p_i\}$, to translate the samples into physical momenta. A uniform distribution can be realized by
the Rambo mapping~\cite{Platzer:2013esa}. However, typically adaptive multi-channel importance-sampling techniques are used to better map out the modes of the integrand, where each channel represents a unique parametrization of the phase space. To efficiently generate initial state momenta, rather than integrating over the Bjorken variables $x_{a/b}$, the quantities
\begin{align}
    \tau=x_ax_b\,,\quad y=\frac12\ln\left(\frac{x_a}{x_b}\right)\,,\quad\text{with}\quad dx_adx_b=d\tau dy\,,
\end{align}
are typically used. While for $y$ a uniform sampling distribution is suitable, $\tau$ is typically chosen according to $p(\tau)\propto 1/\tau$. The samples used to integrate the cross section can be used directly to analyze arbitrary differential distributions, i.e.\ they can be interpreted as individual scattering events. When unweighting against the maximum weight, we end up with unit-weight events distributed according to the fully differential cross section. 

The phase space sampling methods employed in the multi-purpose event generator \Sherpa, i.e.\ its built-in matrix element generators \Amegic and \Comix, are described in Refs.~\cite{Krauss:2001iv,Gleisberg:2008fv}. In Refs.~\cite{Bothmann:2020ywa,Gao:2020zvv}, neural importance sampling techniques have been studied in conjunction with the \Sherpa multi-channel integrators. 

\subsection{Markov Chain Monte Carlo}

Markov Chain Monte Carlo describes a class of algorithms used for sampling from arbitrary probability distributions, particularly when direct sampling is challenging, e.g. if the probability distribution is far from being uniform. It constructs a Markov chain -- a time series of random variables $x_1,\ x_{2}, \dots$ -- where each variable $x_{n}$ depends only on the previous one, i.e.,
\begin{align}
P(x_n|x_1, x_2, \dots x_{n-1}) = P(x_{n}|x_{n-1}) \, .
\end{align}
The chain is designed such that its stationary distribution $\pi(x)$ matches the target distribution $p(x)$. This means that after many iterations, the distribution of the values in the chain approximates the target distribution, i.e.,
\begin{align}
\lim_{n\rightarrow \infty} P(X_n=x) = \pi(x) \propto p(x) \, .
\end{align}
By running the chain for a sufficient number of steps, MCMC generates samples that approximate $p(x)$. Common MCMC methods include the Metropolis--Hastings algorithm~\cite{Metropolis:1953am,Hastings:1970aa} and Gibbs sampling~\cite{4767596}, which are widely used in fields such as Bayesian statistics, machine learning, and physics.

One limitation of MCMC is the occurrence of autocorrelation. Autocorrelation in Markov chains refers to the correlation between a state at a point in time $n$ and a state at a later time $n+k$. Despite the memoryless nature of Markov chains, autocorrelation can still occur because of the transition probabilities that govern the system's behavior. In particular, the Markov chain's dynamics can induce correlation between successive states, especially if the system tends to remain in the same state for several steps. The autocorrelation can be estimated by computing the correlation between the states of the chain at different time steps. The autocorrelation at lag $k$ is computed by dividing the covariance between the states at time $t$ and $t+k$ by the variance of the states. For a chain with $N$ samples, it can be written as
\begin{align}
\rho(k) = \frac{\frac{1}{N-k} \sum_{i=1}^{N-k} \left( x_{i} - \hat{\mu} \right) \left( x_{i+k} - \hat{\mu} \right)}{\frac{1}{N} \sum_{i=1}^{N} \left( x_{i} - \hat{\mu} \right)^{2} } \, ,
\end{align}
where $\hat{\mu}$ is the sample mean of the chain. Knowing the autocorrelation is crucial for estimating the effective sample size (ESS), which determines the number of actually independent samples. The ESS is computed by considering the autocorrelations over various lags. It can be expressed as
\begin{align}
\label{eqn:ess}
\mathrm{ESS} = \frac{1}{1+2 \cdot N \cdot \sum_{k=1}^{L}\rho(k)} \, ,
\end{align}
where $L$ is the maximum lag considered.

Reducing autocorrelation in a Markov chain typically involves speeding up its mixing, increasing state transitions, or introducing more randomness into the process. Techniques like modifying transition probabilities, using more complex MCMC methods, and thinning the chain are common approaches to achieve this.
The algorithms used in this study are implemented in the BAT.jl software package. BAT.jl is a highly efficient and flexible Julia package that facilitates parallel, batched computations. Its emphasis on performance and parallelism makes it ideal for large-scale, computationally intensive tasks in a variety of fields, including physics and data science. BAT.jl features a set of sampling algorithms that can be chosen by the user. This set includes the Metropolis--Hastings algorithm with two different tuning approaches, a variant of Hamiltonian MC, two different nested sampling algorithms as well as simple importance samplers. BAT.jl also provides a benchmark suite for the comparison of Monte Carlo samplers including different metrics and a set of standardized test cases.

\subsection{MCMC for phase space sampling}

The application of MCMC techniques for phase space integration and event generation in high energy physics
has so far received rather little attention. In Ref.~\cite{Kharraziha:1999iw} an initial attempt has been
presented to apply the Metropolis algorithm to integrate multi-particle processes, thereby directly sampling
in the final-state momenta. This work also addressed processes with multiple production modes, necessitating 
a multi-channel integrator. In Ref.~\cite{Kroeninger:2014bwa} a novel method dubbed (MC)$^3$ has been 
introduced that corresponds to an MCMC algorithm with a mixture proposal distribution, such that new phase space 
points at times get drawn from a process specific multi-channel integrator instead of the Metropolis--Hastings
proposal. Sampling has been performed in a unit hypercube, translating to physical four-momenta through the Rambo 
mapping. In Ref.~\cite{Yallup:2022yxe} the potential of nested sampling for exploring the phase space of scattering
processes has been explored, turning out to be very promising. In contrast to traditional MCMC techniques, 
nested sampling is primarily a numerical integration technique~\cite{Ashton:2022grj}, allowing one to evaluate the evidence integral 
in Bayesian inference, i.e. here the total cross section, thereby providing samples of the target function rather 
as a co-product.    

\section{A case study using BAT.jl and \Sherpa}\label{sec:batsh}

\begin{figure}[h]
\includegraphics[width=\textwidth]{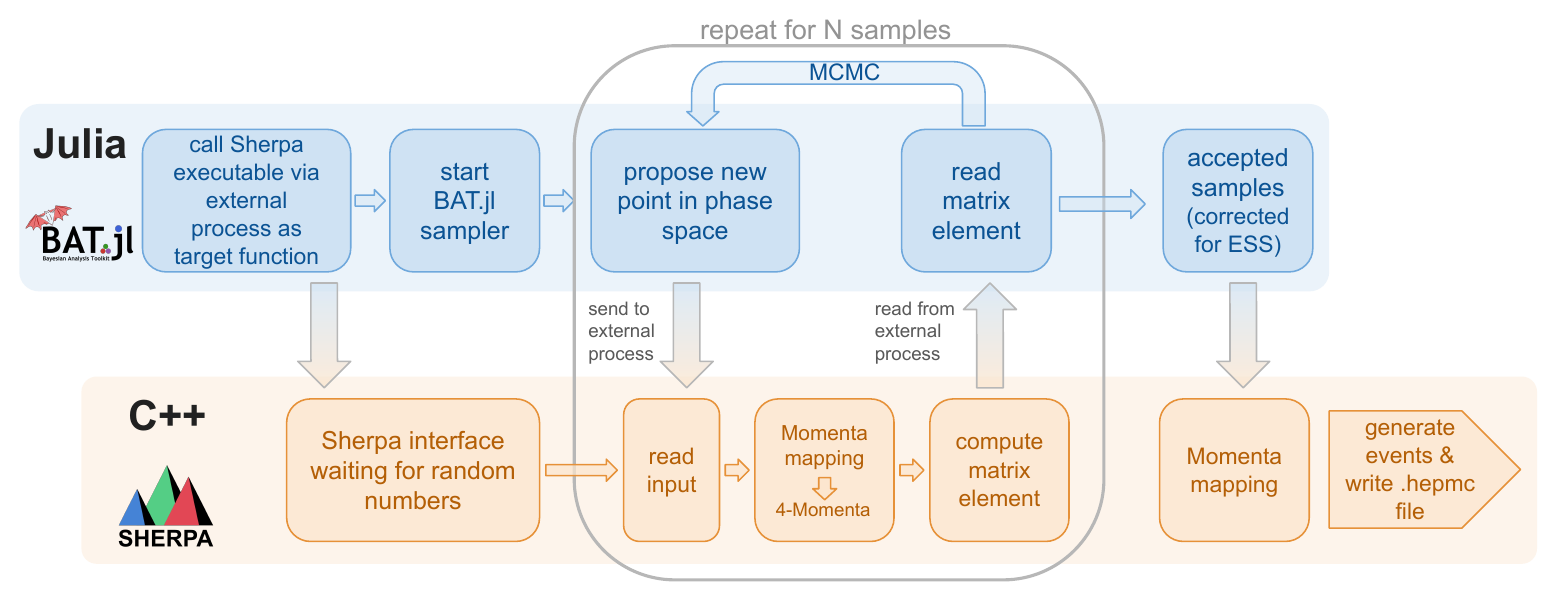}
\caption{Simplified schematic illustration of the interface between BAT.jl and \Sherpa.}
\label{fig:Interface-Diagram}
\end{figure}

To utilize the MCMC sampling algorithms available in BAT.jl for generating MC events according to a given cross section, we have developed an interface connecting the BAT.jl package with the \Sherpa MC generator.
Although our future goal is to fully automate the execution of the MCMC algorithms provided from within \Sherpa, for this initial study, the workflow is initiated manually by running the BAT.jl toolkit, using the desired differential cross section as target function.
The schematic diagram in Fig.~\ref{fig:Interface-Diagram} illustrates the current workflow of the interface.

For the interface, a modified \Sherpa executable is employed. This executable reads the specified \Sherpa configuration file in the \texttt{yaml} format and initializes all internal computations as usual, up to the stage where random numbers are drawn for the generation of the four-momenta, either by using the Rambo or the process-specific multi-channel mappings. At this point, the program halts, waiting for the input of the appropriate number of random variables. Once these are provided, e.g.\ by BAT.jl, the program resumes, and performs the phase-space mapping to the physical four-momenta of the incoming and outgoing particles.
Based on these momenta, the matrix element and phase-space factors are evaluated, and the resulting value, i.e.\ the event weight, is returned.

This whole process is started and controlled from within the Julia code, where the described modified executable is launched inside an external process. The BAT.jl sampling algorithms then pass their next proposed point to the external process and retrieve the computed matrix element value from \Sherpa as the value of the target distribution.
This procedure continues until the desired number of MCMC steps is completed. In the end, the accepted phase space points are passed through the interface a last time to compute the corresponding four-momenta and save the events in \HepMC format~\cite{Buckley:2019xhk}.

To test the interface and investigate the applicability of the approach for a realistic use case, we study $Z+3$ jets production at the LHC, specifically the partonic channel $gg \rightarrow e^+ e^-  d\bar d g$ in $pp$ collisions at $\sqrt{s} = 13\,\mathrm{TeV}$.
We apply a cut on the invariant dilepton mass of $m(\ell\ell)>66\,\mathrm{GeV}$. Jets are reconstructed using the anti-$k_t$ algorithm~\cite{Cacciari:2008gp}, using $R=0.4$ and $p_{T,j}>20\,\mathrm{GeV}$. 

\begin{figure}[h]
\includegraphics[width=0.48\textwidth]{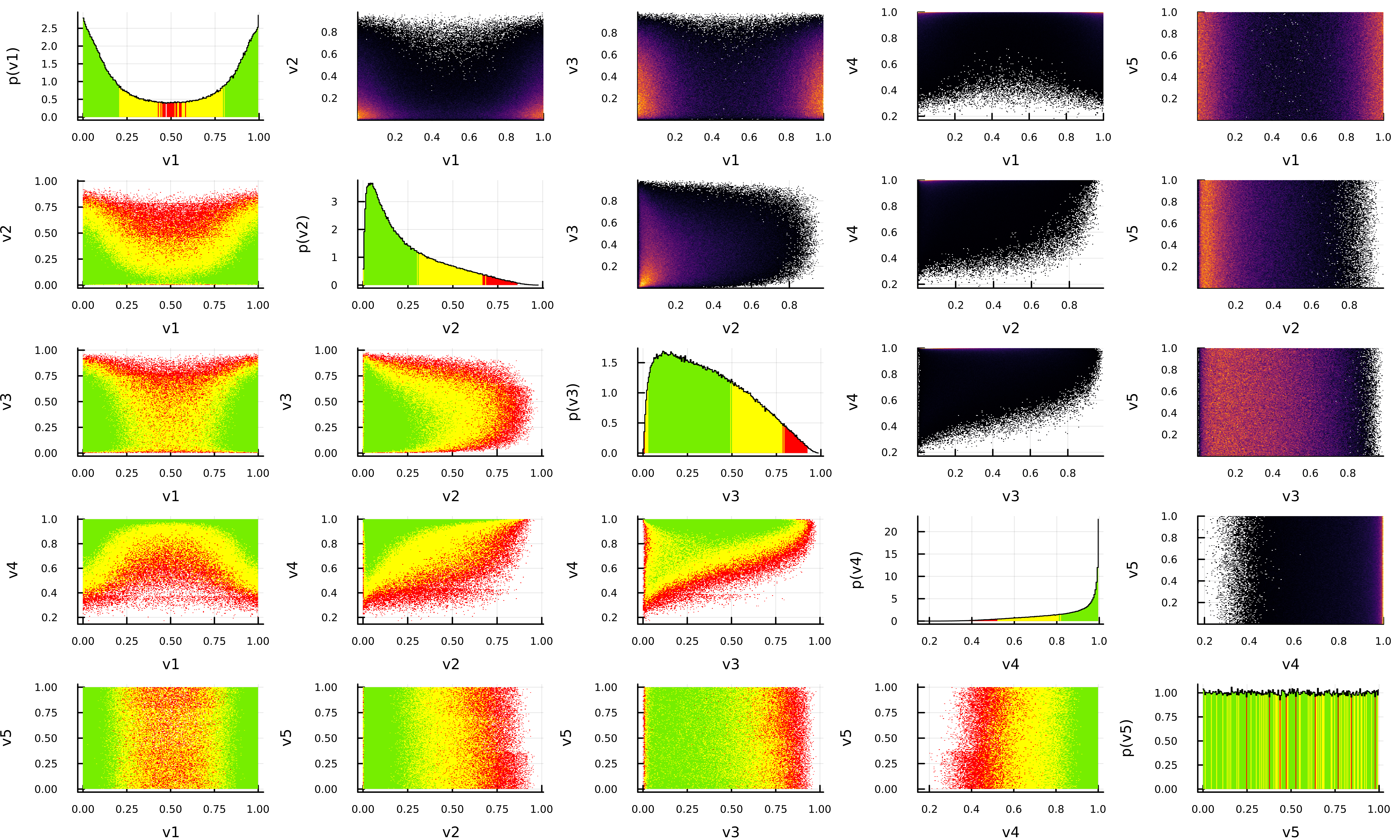}\hfill
\includegraphics[width=0.48\textwidth]{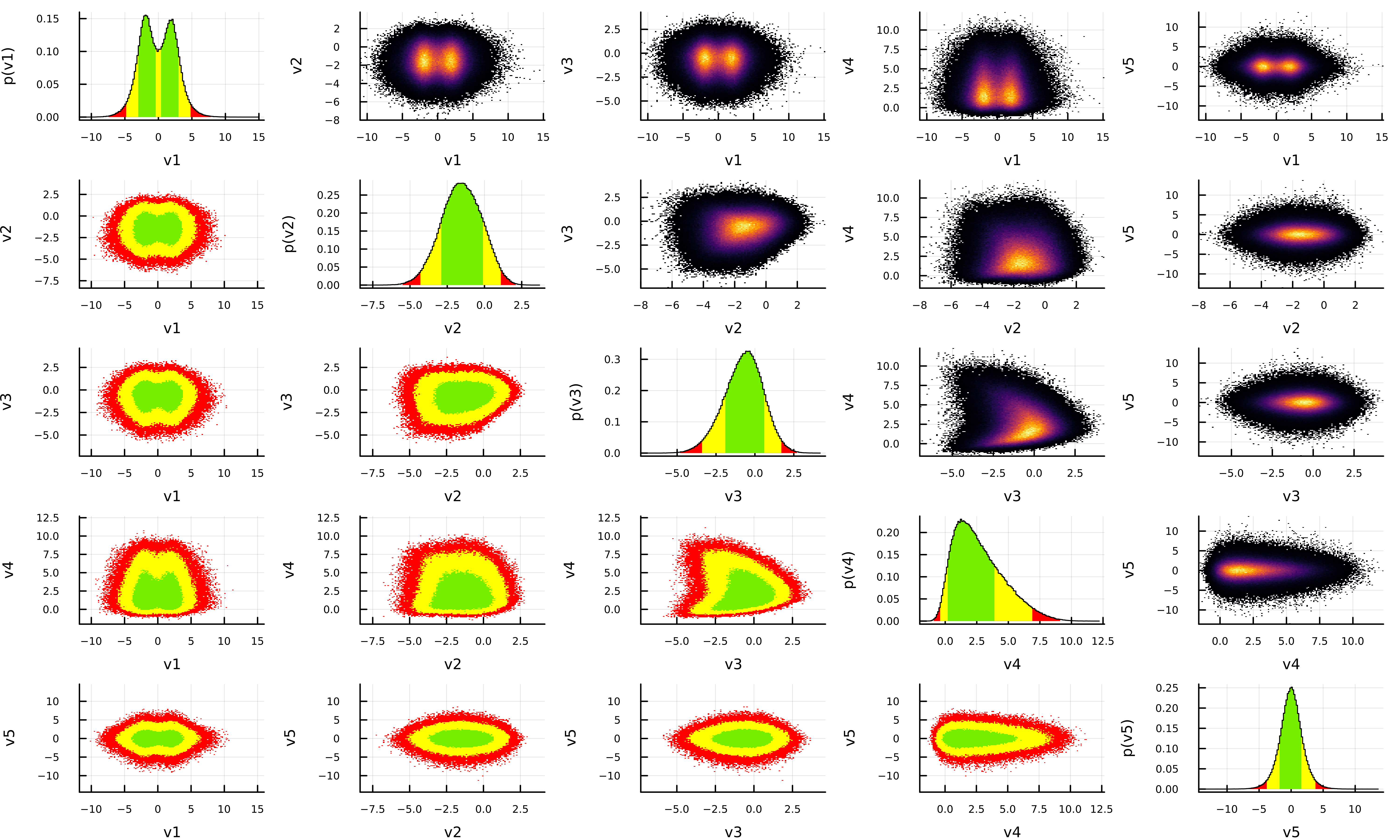}
\caption{One- and two-dimensional marginalized distributions of five of the 11 parameters describing the momenta of the outgoing particles before (left) and after applying the logit transformation (right).}
\label{fig:parameter-space}
\end{figure}

The phase space in this example has 13 dimensions: while two variables represent the longitudinal momentum fractions of the incoming gluons, parametrized by the initial-state variables $\tau$ and $y$, the remaining 11 variables describe the on-shell momenta of the five particles in the final state (after enforcing total four-momentum conservation). 
The adaptive multi-channel importance sampling implemented in \Sherpa results in 96 different final state channels for this specific process. As an example, we show the one- and two-dimensional marginalized distributions of the 13-dimensional phase space for this process in Fig.~\ref{fig:parameter-space} when using the variables $\tau$ and $y$ for the initial state and the specific parametrization of an arbitrary channel describing the momenta of the outgoing particles. We use a Metropolis--Hastings algorithm with RAM tuning \cite{Vihola2012} and the Gelman--Rubin convergence criterion \cite{gelman1992inference} for the sampling.

Most marginalized distributions in Fig.~\ref{fig:parameter-space} (left) are either nearly uniform or have high probabilities concentrated near the boundaries of the allowed parameter ranges, which poses a challenge for MCMC algorithms. To mitigate this, we apply a logit transformation, defined as $q(x) = \log\left(\frac{x}{1-x}\right)$, during the sampling process. This transformation redistributes the probabilities towards the phase space center, making it easier for MCMC algorithms to sample from the transformed distribution. The corresponding marginal distributions are displayed in Fig.~\ref{fig:parameter-space} (right) and show a clear concentration of mass in the central regions. Although the shapes of the distributions differ in detail when using the various channels and their phase space parameterization, any channel or a combination of channels can be used for sampling.

\begin{figure}[h]
\centering
\includegraphics[width=0.66\textwidth]{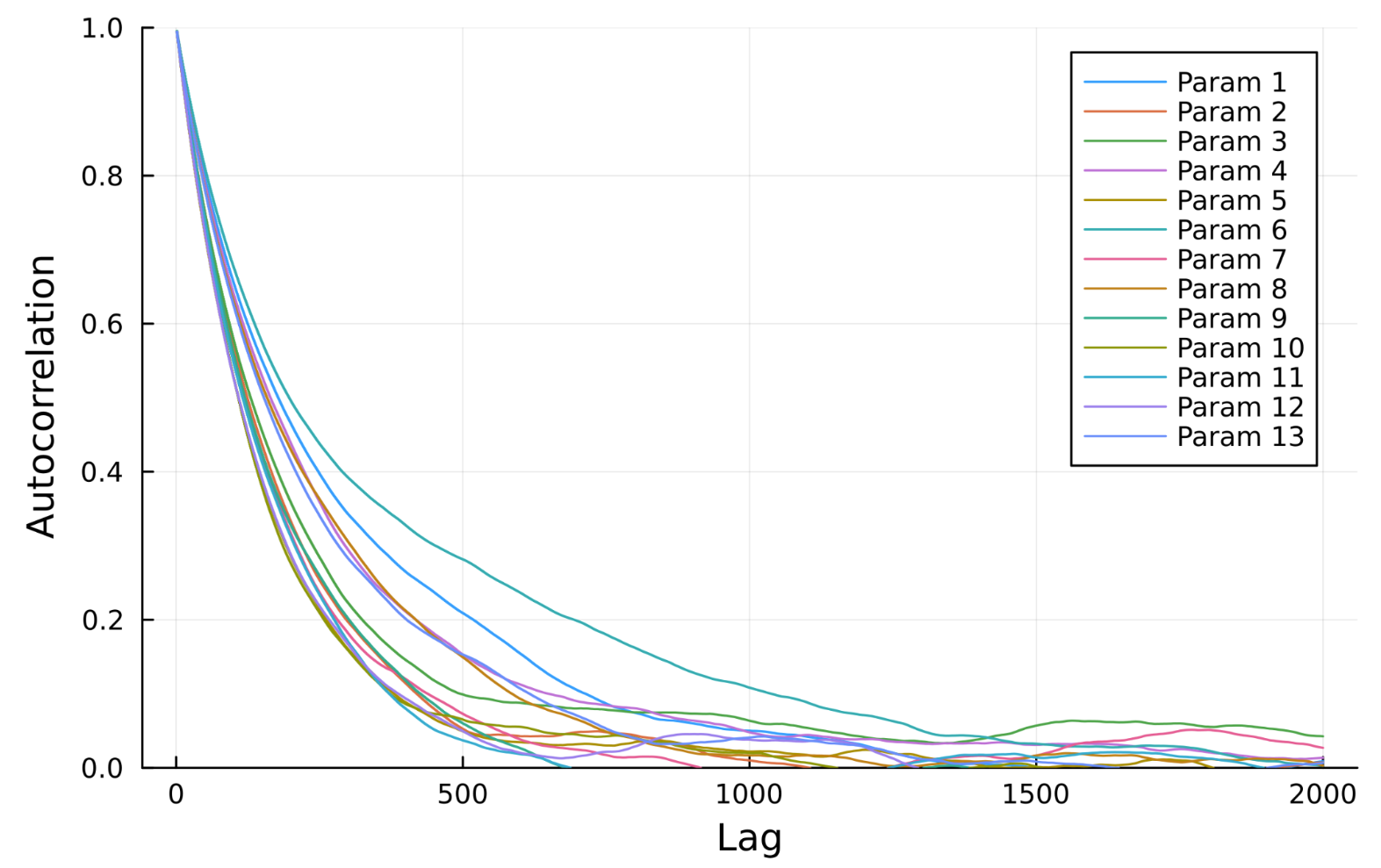}
\caption{Autocorrelation as a function of the lag for the 13 (logit-transformed) parameters using the Metropolis--Hastings algorithm for the $Z+3$ jets example.}
\label{fig:autocorrelation}
\end{figure}

When sampling the logit-transformed phase space with the MCMC algorithms provided in BAT.jl, the tuning of the Markov chains works well and fast convergence is achieved as quantified by the Gelman--Rubin convergence criterion implemented in BAT.jl. %
However, as discussed in Sec.~\ref{sec:methods}, the generated samples for this example are not independent and the autocorrelation between them is typically quite large. 
This can be seen in Fig.~\ref{fig:autocorrelation}, where the autocorrelation for each of the 13 parameters is shown as a function of the lag, i.e., the time interval between the samples. The autocorrelation observed for the chosen channel mapping starts at 1 and drops to a level of 10\% with lags between 300 and 1000, depending on the parameter.
Autocorrelation in general is observed for the Rambo mapping as well as for the individual importance sampling channels. However, the decay rate is slightly different for each channel and parameter.

To account for the observed autocorrelation, we draw a large number of samples until the effective sample size, estimated using Eq.~\eqref{eqn:ess}, reaches a target value. Subsequently, we perform a randomized resampling of all generated samples. When comparing to independent and identically distributed samples obtained from a stand-alone \Sherpa-only run, we make sure that the number of \Sherpa events matches the ESS of the samples provided by the \Sherpa + BAT.jl interface.

\begin{figure}[h]
\includegraphics[width=0.5\textwidth]{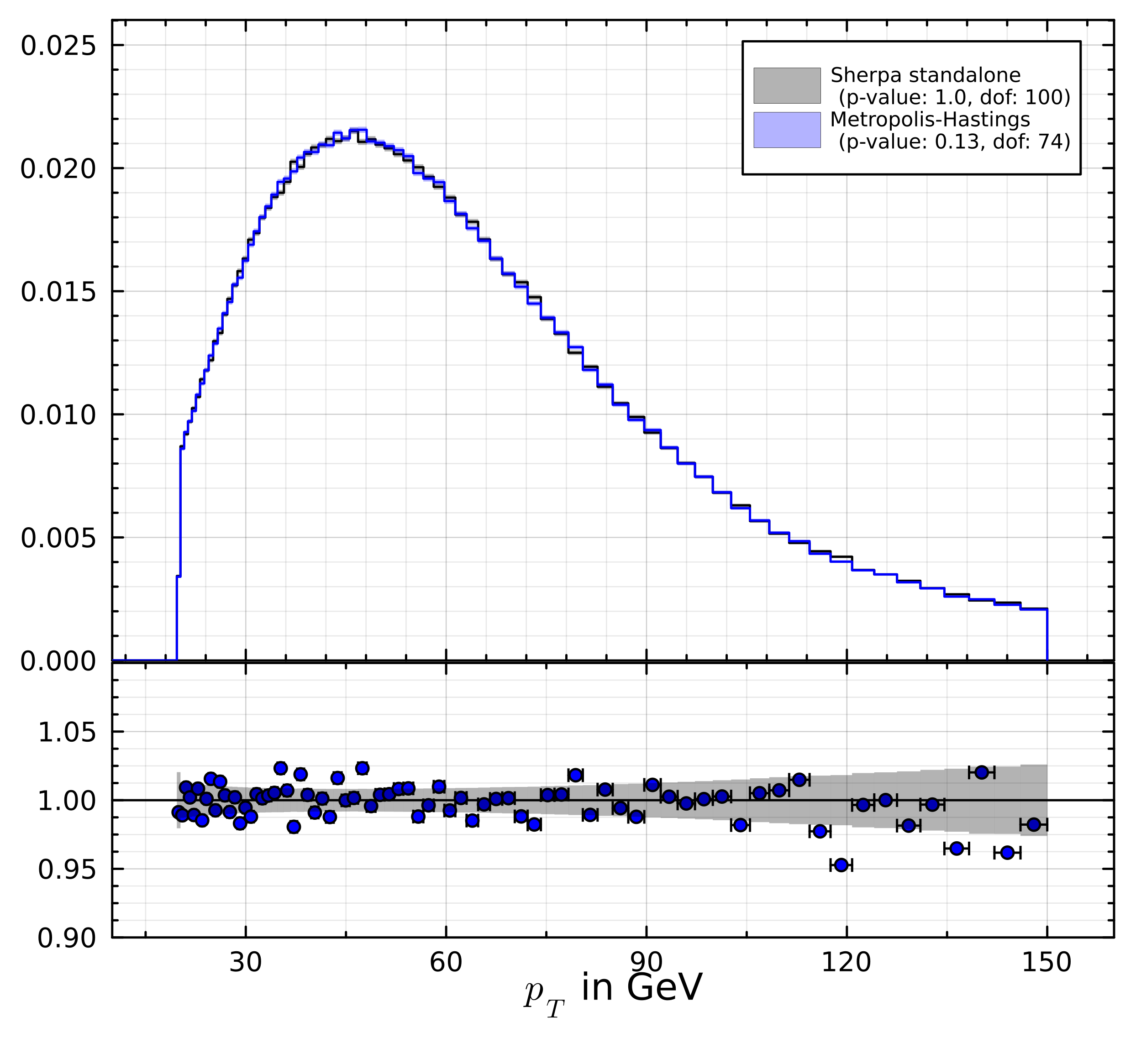}
\includegraphics[width=0.5\textwidth]{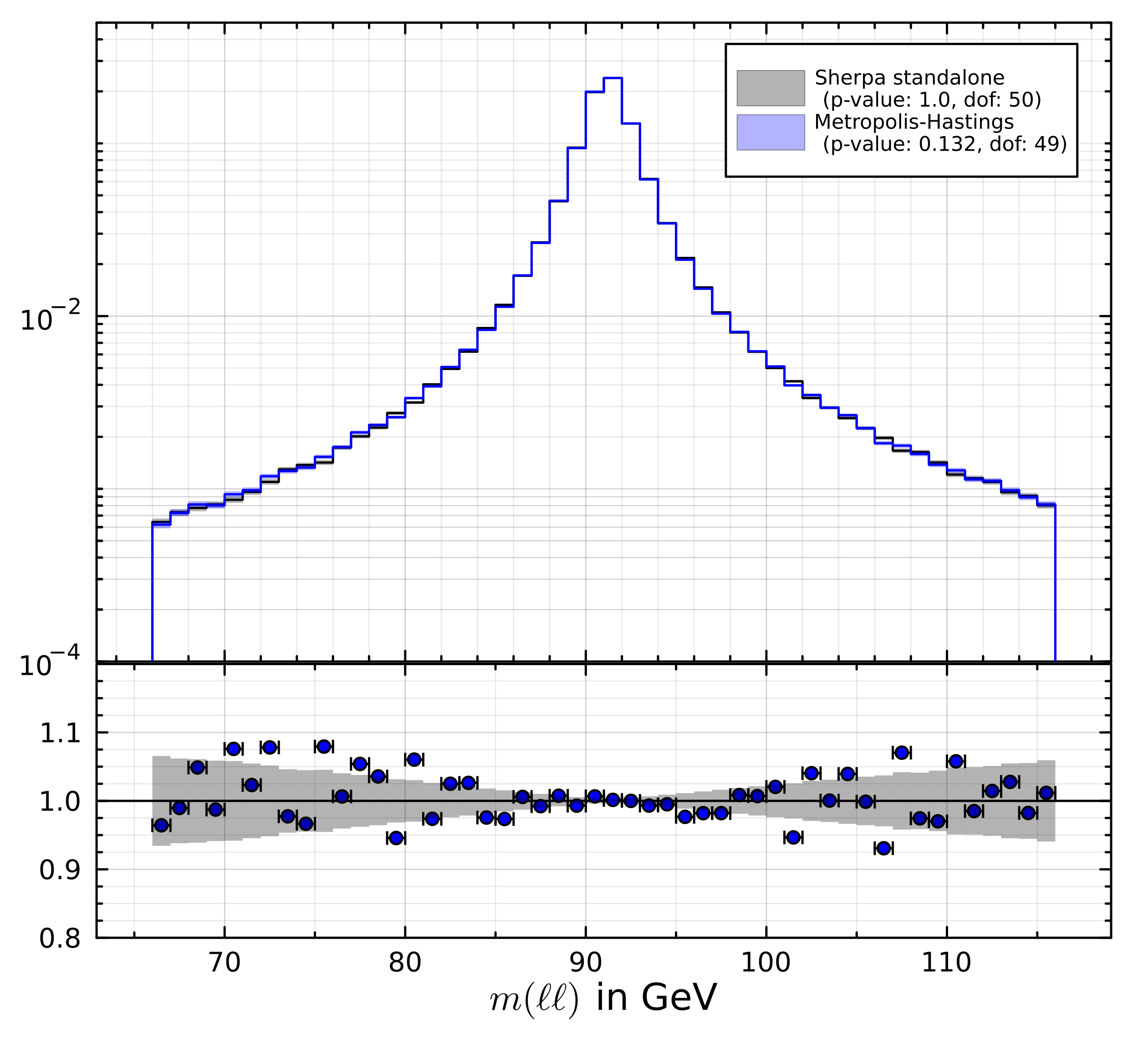}
\caption{Distributions of the leading lepton $p_T$ (left) and the dilepton invariant mass (right) as generated by running \Sherpa standalone (gray) and by using our interface with the Metropolis--Hastings algorithm (blue). The blue markers represent the ratio of the bin values, while the gray band reflects the combined statistical uncertainties from both samples, calculated as the squared sum. The $p$-value, derived from a $\chi^2$-test, quantifies the compatibility of the two distributions.}
\label{fig:observables}
\end{figure}

As an example for such a comparison, Fig.~\ref{fig:observables} shows the distributions of two physical quantities defined in the $Z+3$ jets process, namely the transverse momentum, $p_T$, of the leading lepton (left) and the invariant mass of the two leptons, $m(\ell \ell)$ (right). The samples generated by \Sherpa (gray) and through our interface (blue) are shown to be in good agreement within the statistical uncertainties. The bottom panels show the ratio of each bin and the corresponding combined uncertainty. No bias is observed and the data fluctuations are reflected by the estimated statistical uncertainties. This is also indicated by the $p$-values derived from a classical $\chi^{2}$-test. 

\FloatBarrier

\section{Conclusions and Outlook}
\label{sec:conclusions}

We have developed an interface between the BAT.jl software package and the \Sherpa event generator. This enables the sampling of differential cross sections using a variety of different (MCMC) algorithms. In the physics-relevant case study of $Z+3$ jets production, we have demonstrated that MCMC algorithms can be used to sample from the corresponding phase space and generate distributions of the differential cross sections without bias. While this allows for the use as an event generator, we observe an autocorrelation among the individual samples/events. In future studies, we will investigate different ways to reduce autocorrelation among samples and test the algorithm on more complex final states and processes featuring pronounced resonance structures. We will also quantify the resulting reduction in computing time with the MCMC approach to event generation.

\section*{Acknowledgements}

The authors are supported by the German Federal Ministry of Education and Research
(BMBF) in the ErUM-Data action plan via the KISS consortium (Verbundprojekt 05D2022). TJ expresses his gratitude to CIDAS for supporting him with a fellowship.

\bibliography{references}
\end{document}